\documentclass[conference]{IEEEtran}
\IEEEoverridecommandlockouts
\usepackage{cite}
\usepackage{amsmath,amssymb,amsfonts,amsthm}
\usepackage{algorithm,algorithmic}
\usepackage{array}
\usepackage{graphicx}
\usepackage{textcomp}
\usepackage{xcolor}
\usepackage{booktabs}

\usepackage{subfigure}
\usepackage{bm}

\def\BibTeX{{\rm B\kern-.05em{\sc i\kern-.025em b}\kern-.08em
    T\kern-.1667em\lower.7ex\hbox{E}\kern-.125emX}}
\begin{document}

\title{Peer-to-Peer Energy Cooperation in Building Community over A Lossy Network \\
\thanks{This work was supported in part by National Natural Science Foundation of China (71971183 and 72071100), Guangdong Basic and Applied Basic Research Fund (2019A1515111173), Young Talent Program (Department of Education of Guangdong) (2018KQNCX223) and High-level University Fund (G02236002).
}}

\author{\IEEEauthorblockN{Cheng Lyu\IEEEauthorrefmark{1}\IEEEauthorrefmark{2}, Youwei Jia\IEEEauthorrefmark{2} and Zhao Xu\IEEEauthorrefmark{1}\IEEEauthorrefmark{3}}
	\IEEEauthorblockA{\IEEEauthorrefmark{1}Department of Electrical Engineering, The Hong Kong Polytechnic University, Hong Kong}
	\IEEEauthorblockA{\IEEEauthorrefmark{2} Department of Electrical and Electronic Engineering, Southern University of Science and Technology, Shenzhen, China}
	\IEEEauthorblockA{\IEEEauthorrefmark{3}Shenzhen Research Institute and Research Institute for Smart Energy, The Hong Kong Polytechnic University, Hong Kong\\
	Email: cheng.lyu@connect.polyu.hk, jiayw@sustech.edu.cn, eezhaoxu@polyu.edu.hk}}


\maketitle

\begin{abstract}
Energy management of buildings is of vital importance for the urban low-carbon transition. This paper proposes a sustainable energy cooperation framework for the building community by communication-efficient peer-to-peer transaction. Firstly, the energy cooperation of buildings is formulated as a social welfare maximization problem, in which buildings may directly trade energy with neighbors. In addition, considering privacy concerns and communication losses arisen in peer-to-peer energy trading, a communication-failure-robust distributed algorithm is developed to achieve the optimal energy dispatch solutions. Finally, simulation results show that the proposed framework substantially reduces the total cost of the building community and the algorithm is robust to communication losses in the network when only part of links (even one link) are active during iterations. 
\end{abstract}


\section{Introduction}
Buildings consume more than 40\% of total energy use reported by United Nations Environment Program \cite{b1}, and the building sector in Hong Kong even amazingly accounts for over 93\% of total electricity consumption \cite{b2}. In many countries, conventional buildings are indeed undergoing a transition to smart energy buildings, to enhance the energy efficiency by heating, ventilation, air conditioning (HVAC), energy storage system (ESS), and renewable energies (e.g., wind energy and solar energy). Recently, the concept of shifting from energy management of individual buildings to energy cooperation in the building community becomes popular. Particularly, the development of peer-to-peer technology makes peer-to-peer energy cooperation become realizable and promising in collaboratively managing the local energy while preserving the information privacy \cite{b3}.

The peer-to-peer energy trading framework in smart grids has been investigated in many works. The existing results can be roughly classified into two groups: \textit{centralized models} and \textit{distributed models}. In the first class, the peer-to-peer trading is 
implemented, to some extents, in aid of a coordinator in the network. For example, \cite{b4} proposes a peer-to-peer trading platform for residential houses to coordinate the demand response with renewable energy generations. With the increasing integration of renewables, traditional passive consumers are becoming prosumers that actively manage their production and consumption of energy \cite{b5}. The concept of prosumer preference is introduced in \cite{b5} to distinguish heterogeneous energy sources in the peer-to-peer energy market. In the second class, the information privacy concerns of agents are respected using distributed optimization. All participants in the cooperation aim to seek more incentives in importing or exporting energy. In this regard, \cite{b6} divides the community agents into buyers and seller, and exploits the game theory to model the competition among sellers, the dynamics of buyers, and the interaction between buyers and sellers. Likewise, the non-cooperative Stackelberg game theory is used in \cite{b7} to model the iterative energy trading processes. Moreover, the existence of Stackelberg equilibrium is mathematically proved, under which the social operational cost is minimized. Similar works are presented in \cite{b8} that adopts the cooperative coalition game to enhance the energy use efficiency by directly trading with neighboring agents. In general, the distributed models are reasonable and admirable in implementing the peer-to-peer energy cooperation, since participants are considered as self-interested when engaging in the energy cooperation.

Among the recent literature, the bilateral payment issue is still challenging to resolve in the trading models. To determine a fair scheme to share the cost savings, \cite{b9} proposes a generalized Nash equilibrium framework to model the behaviors of all participants in allocating the profits. In \cite{b10}, Nash bargaining method is employed to achieve a fair payment solution based on market power, defined as the total peer-to-peer trading energy over the horizon. Nevertheless, the detailed energy price at very time slot is rather unclear, as the underlying idea is to allocate the cost savings according to their total contribution. In this paper, an optimal peer-to-peer pricing scheme is proposed using the Lagrangian multipliers with the energy trading balance constraints.

While many works present various models for the peer-to-peer energy trading framework, few of them address the possible communication losses arisen in the trading processes. In fact, peer-to-peer transaction introduces a large burden on communication links due to iterative message exchanges in the negotiation process \cite{b11}. Indeed, the distributed system may suffer from communication failures, on top of the possible asynchronism in agent updates \cite{b12}. In this paper, the authors are also interested in solving the distributed energy cooperation problem in the presence of communication losses. 

Considering the above challenges, this paper proposes a sustainable energy cooperation framework for the building community. The main contributions of this paper are threefold:

\begin{enumerate}
	\item The peer-to-peer energy cooperation framework is proposed for the building community, in which individual buildings can flexibly manage HVAC loads, ESSs, utility grid trading and peer-to-peer trading energy.
	
	\item The peer-to-peer trading price is inherent in the proposed framework using the Lagrangian multipliers associated with the trading energy balance constraint. As such, the peer-to-peer trading price is optimally determined at every time slot.
	
	\item The possible communication losses in the iterative peer-to-peer message exchanges are newly considered and modeled in this paper. In addition, a communication-failure-robust distributed algorithm over the lossy network is proposed for the peer-to-peer energy cooperation problem.
	
\end{enumerate}

The rest of this paper is organized as follows: In Section \ref{systemmodel}, we present the system model of energy buildings. In section \ref{problemformulation}, the problem for energy cooperation is formulated. A fully distributed algorithm considering communication failures is proposed in Section \ref{ALGO}. Simulation results are presented in Section \ref{results}. Finally, Section \ref{conclusion} summarizes this paper.

\section{System Model Buildings}\label{systemmodel}
In this paper, let $\mathcal{N} = \{1,2,\dots,N\}$ represent the buildings in the community; $\mathcal{E} $ denotes the set of all links between buildings; agents $i$ and $j$ are called neighbors if the link $(i,j) \in \mathcal{E}$; $ \mathcal{N}_i$ represents the neighborhood set of $i\in \mathcal{N}$. The scheduling horizon is divided into $ T $ time slots with 1-hour interval and $ \mathcal{T}= \{1,2,\dots,T\}$.

\subsection{Controllable HVAC Load}
HVAC loads, dominating the energy consumption in buildings, are expected to adjust the indoor temperature according to the evolution law \cite{b9}:
\begin{equation}
T_{i,t}^{\rm in}=\left( 1-\frac{1}{J_i R_i}\right)T_{i,t-1}^{\rm in}  + \frac{T_{i,t}^{\rm out}}{J_i R_i} -\frac{\eta_ip^{\rm hvac}_{i,t}}{J_i},\quad \forall t \label{temp}
\end{equation}
\begin{equation}
T_{i,\min}^{\rm in}\le T_{i,t}^{\rm in} \le T_{i,\max}^{\rm in},\quad \forall t \label{temp_range}
\end{equation}
where $T_{i,t}^{\rm in} $ and $T_{i,t}^{\rm out} $ are the indoor and outdoor temperature, $p^{\rm hvac}_{i,t}$ is the energy consumption, $J_i$ is the HVAC parameter, $R_i$ is the envelope parameter, $\eta_i$ is the energy efficiency, $[T_{i,\min}^{\rm in},T_{i,\max}^{\rm in}]$ is the acceptable range, of building $ i \in \mathcal{N}$. Note that, \eqref{temp} is derived based on the RC model with the 1-hour time resolution, to roughly estimated the energy consumption \cite{b14}. 

The discomfort cost of HVAC loads are modeled using the deviation from the desired temperature: 
\begin{equation}
C_{i,t}^{\rm hvac}(p^{\rm hvac}_{i,t})= \beta_i \left( T_{i,t}^{\rm in} - T_{\rm desir}^{\rm in}\right) ^2, \quad \forall t \label{cost_hvac}
\end{equation}
where $ T_{\rm desir}^{\rm in}$ is the desired temperature, $\beta_i$ is the discomfort parameter, of building $ i \in \mathcal{N}$. 

\subsection{Energy Storage System}

Let $SOC_{i,t},p^{\rm c}_{i,t}$ and $p^{\rm d}_{i,t}$ denote the State of Charge (SOC), charging and discharging power of the ESS in the building $i$, respectively. Then, the following constraints capture the operation characteristic of ESS: 
\begin{subequations}\label{soc}
	\begin{equation}
	SOC_{i,t} = SOC_{i,t-1} + \frac{\eta_i^{\rm c}p^{\rm c}_{i,t}}{B_i} - \frac{p^{\rm d}_{i,t}}{\eta_i^{\rm d}B_i} ,\quad \forall t \label{socevo}
	\end{equation}
	\begin{equation}
	SOC_{i,\min} \le SOC_{i,t} \le SOC_{i,\max},\quad \forall t\label{socminmax}
	\end{equation}
	\begin{equation}
	SOC_{i,T}\ge SOC_{i,\rm 0} \label{socend}
	\end{equation}
	\begin{equation}
	0 \le p^{\rm c}_{i,t} \le p^{\rm c}_{i,\max},0 \le p^{\rm d}_{i,t} \le p^{\rm d}_{i,\max},\quad \forall t \label{chmaxmin}
	\end{equation}
\end{subequations}
where $B_i$ denotes the capacity of the ESS, and $\eta_i^{\rm c},\eta_i^{\rm d}$ are charging and discharging efficiencies. Specifically, \eqref{socevo} represents the evolutionary law of SOC; \eqref{socminmax} suggests the permissible operational SOC range; in \eqref{socend}, the SOC at the end of the horizon is no less than the initial SOC; \eqref{chmaxmin} suggests that the charging and discharging power is subject to maximal power limits. 

Due to the degradation in frequent charging and discharging processes, the operational cost of ESS is described as follows:
\begin{equation}
C_{i,t}^{\rm ess}(p^{\rm c}_{i,t},p^{\rm d}_{i,t}) =  \lambda_i^{\rm c} p^{\rm c}_{i,t}+\lambda_i^{\rm d}  p^{\rm d}_{i,t}  , \quad \forall t\label{besscost}
\end{equation}
where $\lambda_i^{\rm c}$ and $\lambda_i^{\rm d}$ are the unit cost of charging and discharging. 

\subsection{Utility Grid Energy Trading}
In the building community, energy buildings can directly purchase energy from the utility company when the energy deficit occurs; when the total energy supply exceeds the local demand, the buildings can sell the surplus energy to the utility.
Let $p^{\rm b}_{i,t}\ge 0$ denote the quantity of energy purchased from the utility company and $p^{\rm s}_{i,t}\ge 0 $ denote the energy sold to the utility by the building $i \in \mathcal{N}$, at time slot $t$. As such, the utility grid energy trading cost is formulated as follows:
\begin{equation}
C_{i,t}^{\rm grid} (p^{\rm b}_{i,t},p^{\rm s}_{i,t}) =\mu_t^{\rm b} p^{\rm b}_{i,t}  - \mu_t^{\rm s}p^{\rm s}_{i,t} ,\quad \forall t
\end{equation}
where $\mu_t^{\rm b}$ denotes the energy buying price, and $\mu_t^{\rm s}$ denotes the selling price. In general, the buying price is greater than the selling price, $\mu_t^{\rm b} \ge \mu_t^{\rm s}$. 
\section{Energy Cooperation Problem Formulation }\label{problemformulation}

In this section, we consider a bilateral energy trading between building $i\in \mathcal{N}$ and $j \in \mathcal{N}_ i$, and let $e_{i,t}^j$ represent the energy exchange quantity from building $j$ to building $i$ at time slot $t$. 
Due to the constructional and physical limit, the energy exchange amount limit and power balance should be satisfied:
\begin{equation}
-p^{\rm line}_i \le p^{\rm b}_{i,t}- p^{\rm s}_{i,t}  \le  p^{\rm line}_i, 
-p^{\rm p2p}_i \le e_{i,t}^j \le  p^{\rm p2p}_i, \quad \forall t  \label{buysellminmax2}
\end{equation}
\begin{equation}
p^{\rm b}_{i,t} - p^{\rm s}_{i,t} + p^{\rm r}_{i,t} + \sum_{j \in \mathcal{N}_ i}e_{i,t}^j= p^{\rm load}_{i,t} + p^{\rm hvac}_{i,t} + p^{\rm c}_{i,t}-p^{\rm d}_{i,t} , \quad \forall t \label{newbalance}
\end{equation} 
where $p^{\rm line}_i$ and $p^{\rm p2p}_i$  represents the power limit of distribution line and p2p line, $p^{\rm r}_{i,t}$ represents the renewable energy output, $p^{\rm load}_{i,t}$ represents the basic uncontrollable load, in the building $i \in \mathcal{N}$.

For the bilateral peer-to-peer energy trading between building $i$ and $j$, the following energy trading and price balance should be satisfied:
\begin{equation}
e_{i,t}^j + e_{j,t}^i = 0, \quad \forall i \in \mathcal{N},\forall j \in \mathcal{N}_ i, \forall t  \label{eijequ}
\end{equation}
\begin{equation}
\pi_{i,t}^j = \pi_{j,t}^i={\pi}_{(i,j),t}, \quad \forall (i,j) \in \mathcal{E},\forall t \label{policy}
\end{equation}
where $\pi_{i,t}^j$ denote the energy price of building $i$ set for $j$, and $\pi_{(i,j),t}$ represents the dual variable of \eqref{eijequ}.

\textit{\textbf{Remark}: The two key issues to a fair and successful peer-to-peer trading contract are energy equality and payment equality, which are modeled by equations \eqref{eijequ} and \eqref{policy}. These two requirements are widely accepted by the energy sharing mechanism design, as in \cite{b7,b8,b9,b10}.
Note that, the coupling constraint \eqref{eijequ} contains totally $|\mathcal{E}|$ equations over the network. Hence, \eqref{eijequ} can be rewritten in a compact form:
\begin{equation}
\sum\nolimits_{i \in \mathcal{N}} \mathbf{M}_i\mathbf{e}_{i,t}=\textbf{0}, \quad \forall t \label{Eij}
\end{equation}
where $\mathbf{e}_{i,t}=\{e_{i,t}^j\}_{j\in \mathcal{N}_i} \in \mathbb{R}^{N-1}$ collects the neighboring trading energy vector, $\mathbf{M}_i \in \mathbb{R}^{|\mathcal{E}|\times(N-1)}$ is a mapping matrix from nodes to lines, indicating the connection relationship. For a specific building community, $\mathbf{M}_i$ is considered as a known sparse and fixed matrix with elements 1 and 0.}

For the building community with energy cooperation, the social cost minimization problem can be expressed as the sum of all participant buildings over the scheduling horizon $\mathcal{T}$, which is termed as \textbf{P1}:
\begin{equation*}
\begin{aligned}
\text{minimize} &\quad \sum_{i \in \mathcal{N}} \sum_{t \in \mathcal{T}} C^{\rm int}_{i,t}   \\
\text{subject to} &\quad \eqref{temp},\eqref{temp_range},\eqref{soc}, \eqref{buysellminmax2},\eqref{newbalance},\eqref{Eij} 
\\\text{over:}  &\quad \{\mathbf{x}_{i,t},\mathbf{e}_{i,t} \}_{t\in\mathcal{T},i\in\mathcal{N}},\{\bm{\pi}_{t} \}_{t\in\mathcal{T}}
\end{aligned}
\end{equation*}
The decision variables in each building are divided into two groups: local variables $\mathbf{x}_{i,t}=\{p^{\rm hvac}_{i,t},p^{\rm c}_{i,t},p^{\rm d}_{i,t},p^{\rm b}_{i,t},p^{\rm s}_{i,t}\}$, and coupled variables $\mathbf{e}_{i,t} ,\bm{\pi}_{t}$, where $\bm{\pi}_t=\{\pi_{(i,j),t}\}_{ (i,j) \in \mathcal{E}} \in \mathbb{R}^{|\mathcal{E}|} $, ${C}^{\rm int}_{i,t}=C_{i,t}^{\rm hvac}+C_{i,t}^{\rm ess}+C_{i,t}^{\rm grid}$ represents the internal energy cost of building $i$, which is treated as the private information. 

\section{Proposed Distributed Algorithm}\label{ALGO}
\subsection{Distributed Problem Formulation}
Although \textbf{P1} can be solved by centralized optimization tools, this paper is aimed to design a fully distributed framework for privacy preservation. Observing the structure of \eqref{Eij}, one can rewrite \eqref{policy} in a compact form with the aid of $\mathbf{M}_i$:	
\begin{equation}
\mathbf{M}_i ^\mathrm{T} \bm\pi_{t}= \bm \pi_{i,t} , \quad \forall i \in \mathcal{N} , \forall t \label{lamdat}
\end{equation} 

Hence, \textbf{P1} is split into $N$ sub-problems to be distributedly solved by building $i \in \mathcal{N}$, which is termed as \textbf{P2}:
\begin{equation*}
\begin{aligned}
\text{minimize} &\quad   \sum_{t \in \mathcal{T}}\left(  {C}^{\rm int}_{i,t}+  \bm\pi_{t} ^\mathrm{T} \mathbf{M}_i \mathbf{e}_{i,t} \right) \\
\text{subject to} & \quad \mathcal{X}_i=\{\eqref{temp}, \eqref{temp_range},\eqref{soc}, \eqref{buysellminmax2}, \eqref{newbalance}\} \\
\text{over:}  &\quad \{\mathbf{x}_{i,t},\mathbf{e}_{i,t} \}_{t\in\mathcal{T},i\in\mathcal{N}},\{\bm{\pi}_{t} \}_{t\in\mathcal{T}}
\end{aligned}
\end{equation*}

Compared to \textbf{P1}, \textbf{P2} replaces the coupled constraint \eqref{Eij} by the KKT condition in the objective function. Note that the convexity ensures the optimality and equivalence of the reformulation; the second term of the objective function represents the peer-to-peer payments. In doing so, buildings only have to solve the convex optimization problem \textbf{P2} on the local constraint set $\mathcal{X}_i$. However, the existence of the global dual variable $\bm{\pi}_{t}$ suggests that a coordinator is still needed in solving the problem. 

\subsection{DC-ADMM Algorithm}
To efficiently resolve the aforementioned issue in \textbf{P2}, a fully distributed algorithm is devised based on dual-consensus alternating direction method of multipliers (DC-ADMM), which is developed by applying ADMM on the Lagrangian dual problem. In principle, DC-ADMM contains the following steps, at iteration $k=1,2,\dots$, for agents $i\in\mathcal{N}$ \cite{b11}:
\begin{equation}\label{xeupdate}
{\small
	\begin{aligned}
\{\mathbf{x}^k_{i,t},  \mathbf{e}^k_{i,t} \}_{t \in \mathcal{T}}& = {\rm arg}\min_{\mathcal{X}_i} \sum_{t \in \mathcal{T}} \left[ {C}^{\rm int}_{i,t}+\frac{c}{4|\mathcal{N}_i|} \left\|  \frac{1}{c}\mathbf{M}_i\mathbf{e}_{i,t} \right .\right.\\
& \left. \left.- \frac{1}{c}  \mathbf{z}^{k-1}_{i,t}+2\sum\nolimits_{j \in \mathcal{N}_ i}\mathbf{v}_{ij,t}^{k-1}	  \right\|_2^2 \right]  
\end{aligned}}
\end{equation}
\begin{equation}\label{lambdaupdate}
{\small
\bm\pi^{(i),k}_t= \frac{1}{2|\mathcal{N}_i|} \left( 2\sum_{j \in \mathcal{N}_ i}\mathbf{v}_{ij,t}^{k-1} -  \frac{1}{c} \mathbf{z}^{k-1}_{i,t}+\frac{1}{c}\mathbf{M}_i \mathbf{e}^k_{i,t} \right)}
\end{equation}
\begin{equation}\label{vupdate}
\mathbf{v}_{ij,t}^{k}=   \frac{{\bm\pi}^{(i),k}_t+ 
{\bm\pi} ^ {(j),k}_t}{2}, \forall {j \in \mathcal{N}_ i}
\end{equation}
\begin{equation}\label{zupdate}
\mathbf{z}^k_{i,t} =\mathbf{z}^{k-1}_{i,t}+ 2c \sum_{j \in \mathcal{N}_ i} \left( {\bm\pi}^{(i),k}_t -\mathbf{v}_{ij,t}^{k}\right)
\end{equation}

In the updates, $c> 0$ is the given penalty parameter, $\bm\pi^{(i)}_{t}$ collects the local estimate of the global price $\bm\pi_{t}$ in agent $i$, $\mathbf{v}_{ij,t}$ and $\mathbf{z}_{i,t}$ are auxiliary variables. It is noticed that the updates in \eqref{xeupdate} involves a convex optimization problem, and the updates in \eqref{lambdaupdate}\textbf{-}\eqref{zupdate} are simple algebraic operations, at every time slot $t \in \mathcal{T}$.

The DC-ADMM enjoys several desirable features:
Firstly, the aforementioned updates are fully parallel and distributed, except that, in \eqref{vupdate}, every agent $i$ has to exchange the up-to-date ${\bm\pi}^{(i)}_t$ to its neighbors. Secondly, each agent $ i $ uses only local information, i.e., $ {C}^{\rm int}_{i,t} $, $\mathbf{M}_i$, and $ \mathbf{e}^k_{i,t} $. More importantly, the DC-ADMM can be implemented in a lossy network, e.g., communication links between agents may fail at iterations, which motivates us to propose the following algorithm.

\subsection{Communication-Failure-Robust Distributed Algorithm}
\begin{figure}[!t]
	\centering	
	\includegraphics[width=0.35\textwidth,trim=0cm 1cm 5cm 8cm ]{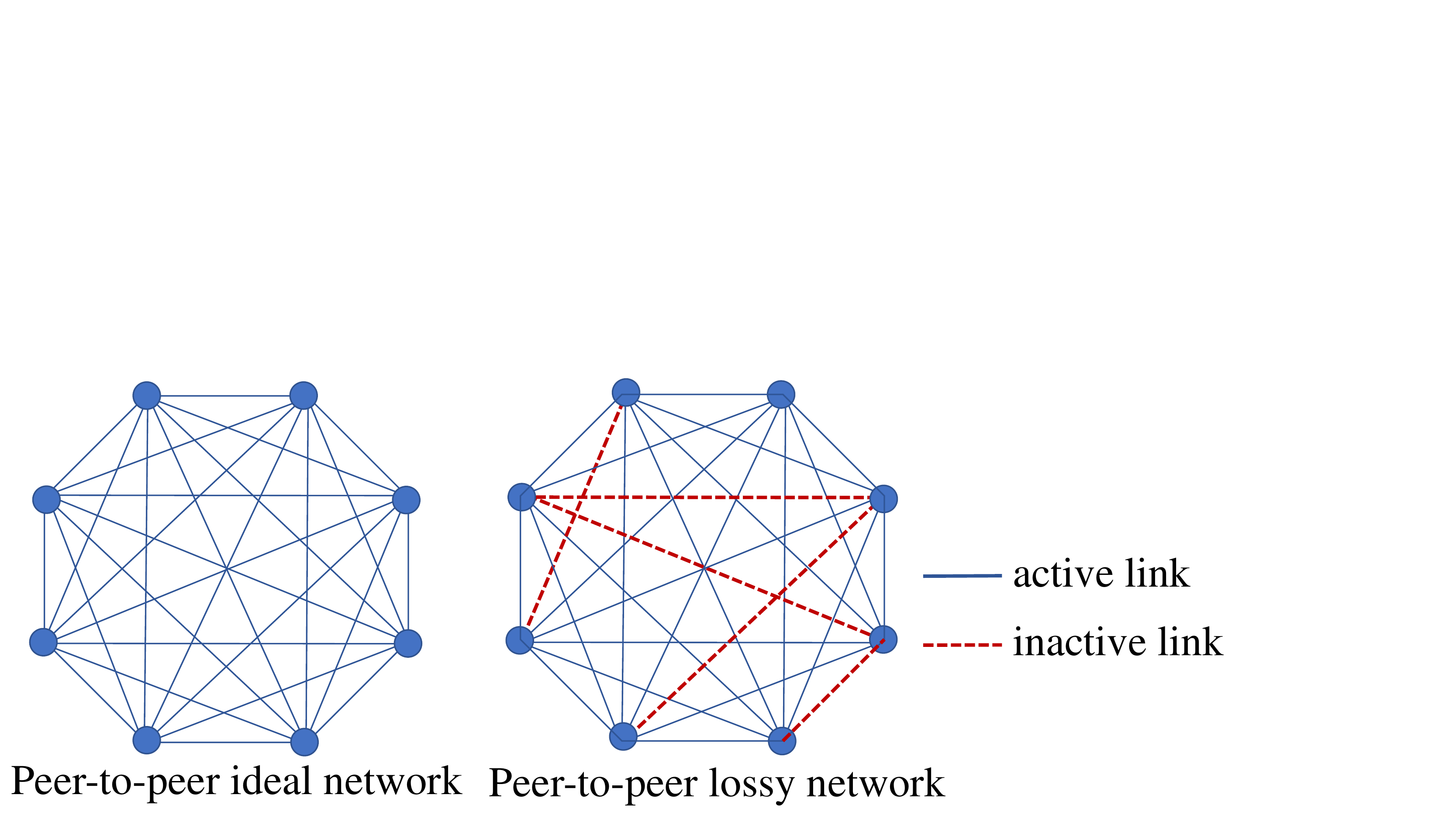}
	\caption{Illustration of the peer-to-peer lossy network: at each iteration, the communication links have a probability to be inactive. }
	\label{network}	
\end{figure}

It is noticed that frequent message exchanges in \eqref{vupdate} may be impractical in imperfect communication networks. To model such a lossy network, we assume that, for each communication link $(i,j)$, there is a probability $\xi_{ij}\in (0,1]$ that the message exchange between agent $i$ and $j$ fails. In such a case, the link is called inactive at this iteration, as shown in Fig. \ref{network}. 

At each iteration $k$, let $\Phi^k \subseteq \mathcal{E} $ be the set of all active communication links, on which the message exchange is successful. Otherwise, the corresponding agents of the link would not update $\mathbf{v}^k_{ij,t}$ as in \eqref{vupdate}, but keep it unchanged: $\mathbf{v}^k_{ij,t}=\mathbf{v}^{k-1}_{ij,t}$.  Likewise, $ \mathbf{z}^k_{i,t} $ is updated using only $\mathbf{v}^k_{ij,t}$ on $(i,j)\in\Phi^k $. Algorithm \ref{algo} summarizes the proposed communication-failure-robust distributed algorithm based on DC-ADMM. 

\begin{algorithm}[h]\begin{small}
	\caption{Communication-Failure-Robust Distributed Algorithm for Peer-to-peer Energy Cooperation}
	\begin{algorithmic}[1]
		\STATE Initialize $\mathbf{x}^0_{i,t},\mathbf{e}^0_{i,t},\bm\pi^{(i),0}_{t}, \mathbf {z}^0_{i,t}, \mathbf{v}^0_{ij,t}=\textbf{0}$,
		 for each building $i \in \mathcal{N}$, $\forall t$. Set iteration $k=1$;
		\REPEAT
		\FOR{all $i \in \mathcal{N}$ (\textit{in parallel})}	
		\STATE Update $ \mathbf{x}^k_{i,t},\mathbf{e}^k_{i,t} $ according to \eqref{xeupdate};
		\STATE Update $ \bm\pi^{(i),k}_{t}$ according to \eqref{lambdaupdate};
		\STATE Exchange $ \bm\pi^{(i),k}_{t}$ with neighbors $j \in \mathcal{N}_ i$;
		\STATE Receive $ \bm\pi^{(j),k}_{t}$, $j\in \{j|(i,j)\in\Phi^k  \}$;
		\STATE Update $\mathbf{v}^k_{ij,t}$ according to \eqref{vupdate} if $(i,j)\in\Phi^k $; otherwise $ \mathbf{v}^k_{ij,t}=\mathbf{v}^{k-1}_{ij,t} $;
		\STATE Update 
		\small $ \mathbf{z}^k_{i,t} =\mathbf{z}^{k-1}_{i,t}+ 2c \sum\nolimits_{j|(i,j)\in\Phi^k } \left( {\bm\pi}^{(i),k}_t -\mathbf{v}_{ij,t}^{k}\right)$;		
		\ENDFOR
		\STATE Set $k=k+1$;
		\UNTIL a predefined stopping criterion is satisfied.
	\end{algorithmic}\label{algo}\end{small}

\end{algorithm}

\section{Case Study}\label{results}
\subsection{Simulation Data Setup}
The considered building community is composed of four energy buildings that are connected with each other and installed with following units besides the basic load from \cite{b13}: 
\begin{enumerate}
	\item Building 1: HVAC load+Solar panel+ESS;
	\item Building 2: HVAC load+Solar panel;
	\item Building 3: HVAC load+ESS;
	\item Building 4: HVAC load+Solar panel+ESS.
\end{enumerate}

The building element parameters are given as: for HVAC loads, $J_i$=1.5kWh/$^\circ$C, $R_i$=1.33$^\circ$C/kWh, $\eta_i$=0.15, and $\beta_i$=$\{2, 2.2, 2.4, 2.3\}\$$/($^\circ$C)$^2$;
for ESSs, $B_i$=200kWh, minimal and maximal SOC are 0.1 and 1, maximal power limit is 50kW, $\eta^{\rm c}_i$=$\eta^{\rm d}_i$=0.95, initial and terminal SOC is 0.5, and $\lambda_i^{\rm c}$=$\lambda_i^{\rm d}$=0.2\$/kWh; $p^{\rm line}_i$=100kW,  $p^{\rm p2p}_i$=50kW,$\mu_t^{\rm b} $=2$\mu_t^{\rm s}$= 1\$/kWh, all costs are in HK dollars.

\begin{figure}[!t]
	\centering	
	\includegraphics[width=0.32\textwidth,trim=0cm 0.5cm 0 0]{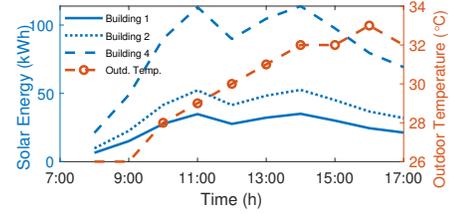}
	\caption{Solar energy and outdoor temperature of the building community.}
	\label{pvtemp}	
\end{figure}

The scheduling horizon is consecutive 10 hours, in which the predicted solar energy and outdoor temperature are shown in Fig. \ref{pvtemp}, $T_{i,\min}^{\rm in}$=20$^\circ$C,  $T_{i,\max}^{\rm in}$=27$^\circ$C, and  $T_{\rm desir}^{\rm in}$=23.5$^\circ$C. In the algorithm, default $\xi_{ij}$=0.2, $c$=4.5 and stopping criterion is iteration to 100. All simulations are implemented using MATLAB2019b, and agent local updates are solved by CPLEX12.9.

\subsection{Performance Evaluation}

\begin{figure}[!t]
	\centering
	\subfigure[]{	
		\includegraphics[width=0.23\textwidth]{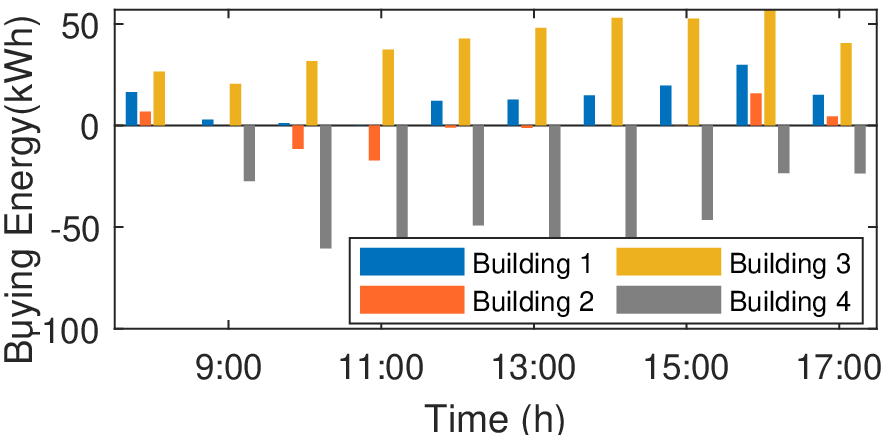}}
	\subfigure[]{
		\includegraphics[width=0.23\textwidth]{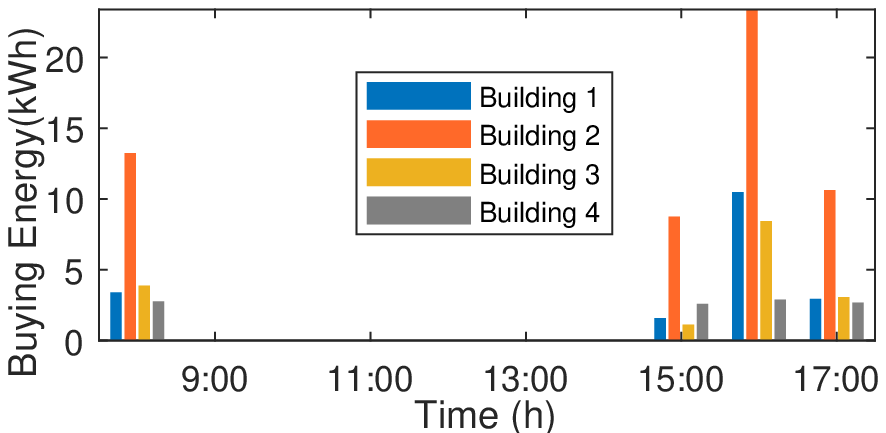}}	\caption{Utility trading energy before and after energy cooperation. The proposed energy cooperation framework can reach consensus utility trading behaviors of  buildings.}\label{plotbuy}	
\end{figure}
\paragraph{Utility energy trading}
Fig. \ref{plotbuy} shows the buying energy from the utility of four buildings before and after cooperation. Before the energy cooperation, all buildings independently conduct energy trading with the utility to realize the local energy balance.
Particularly, building 1 and 3 are energy-deficit since they have to buy energy, while building 4 is relatively energy-sufficient as it sells surplus energy over the scheduling horizon, as seen in Fig. \ref{plotbuy}(a). However, this is not the case after the energy cooperation. In Fig. \ref{plotbuy}(b), four building tends to behave similarly after energy cooperation, in terms of buying energy from the utility grid.
In particular, all buildings buy energy at the first hour and the last three hours. 
This observation demonstrates that the proposed energy cooperation framework can reach a trading consensus with the utility grid.

\begin{figure}[!t]
		\centering	
		\subfigure[]{
			\includegraphics[width=0.26\textwidth,trim=0 0cm 0 0]{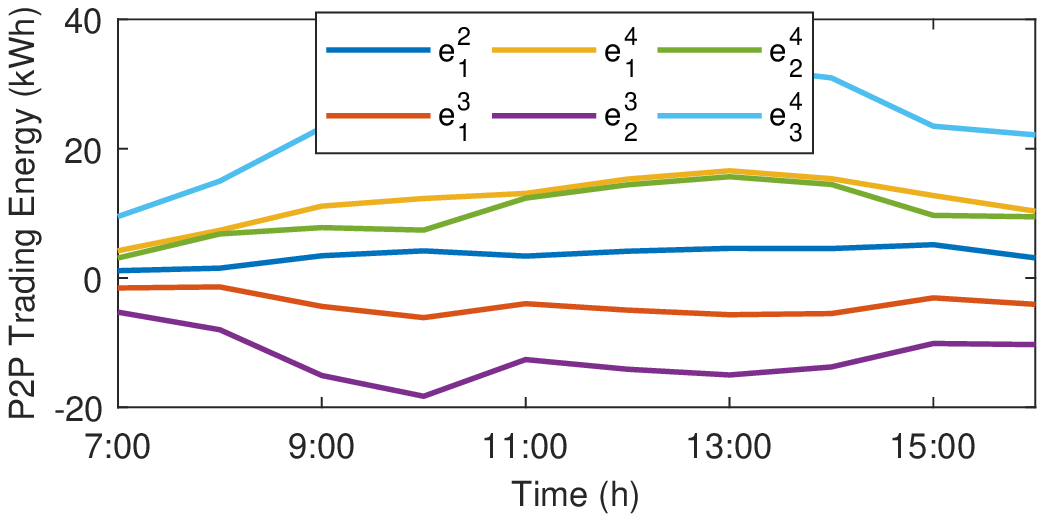}}	
		\subfigure[]
		{\includegraphics[width=0.19\textwidth,trim=0 0cm 0 0]{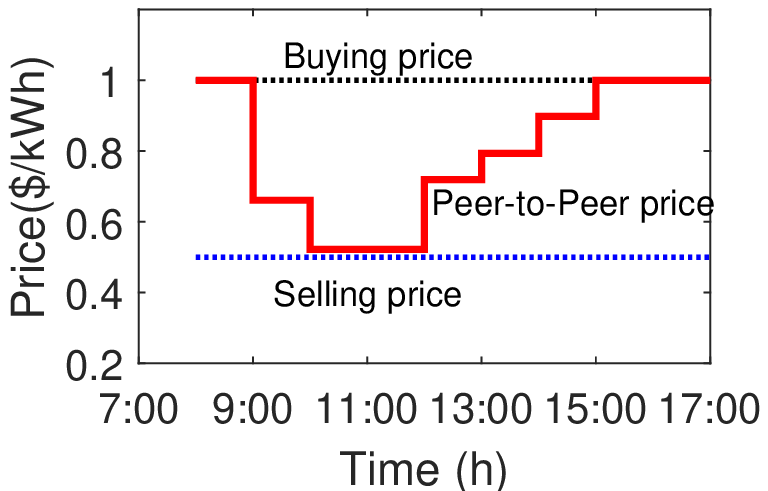}}	
		\caption{Peer-to-peer energy cooperation results. (a) Peer-to-peer energy profiles; (b) peer-to-peer energy trading price.} \label{p2penergy}
\end{figure}

\paragraph{Peer-to-peer energy trading} Fig. \ref{p2penergy}(a) illustrates the peer-to-peer energy profiles of peers after energy cooperation. It is seen that buildings conduct energy trading proactively over the scheduling horizon. Comparing with the utility trading profiles in Fig. \ref{plotbuy}(b), one can also see that, buildings prefer neighbors to realize local energy balance after energy cooperation. In addition, Fig. \ref{p2penergy}(b) shows the peer-to-peer energy trading price, at every time slot over the horizon. Note that, the proposed algorithm can reach a consensus price, even though the price is negotiated in a peer-to-peer manner. Moreover, the price locates between the buying price and selling price, which implies that the cooperation is much more incentive than utility trading.

\begin{figure}[!t]
	\centering	
	\includegraphics[width=0.22\textwidth,trim=0 0.5cm 0 0]{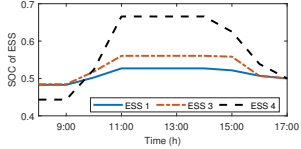}
	\caption{SOC trajectories of ESSs. The proposed energy cooperation framework can desirably harmonize distributed ESSs in the building community.}\label{SOC}	
\end{figure}

\paragraph{ESS usage} Fig. \ref{SOC} depicts the SOC trajectories of ESSs in the building community. It is seen that ESS usage profiles also become almost identical after energy cooperation. This is resulted from the global consensus achieved by iterative peer-to-peer negotiations in the algorithm. This observation suggests that the proposed energy cooperation framework can desirably harmonize the distributed ESSs in the building community.

\paragraph{Cost reduction analysis} Table \ref{tabsocial} summarizes the energy cost of the building community before and after energy cooperation. It is observed that the energy costs of four buildings are all reduced when they participate in the energy cooperation framework. That is, energy cooperation is beneficial to all buildings, especially for those with high energy deficit (e.g., building 3) or high energy surplus (e.g., building 4). This is because that trading energy directly with neighbors, under the peer-to-peer price in Fig. \ref{p2penergy}(b), is more economic than trading with the utility company.   
\begin{table}[h]
	\caption{Energy Cost of Building Community Before and After Energy Cooperation (EC)} 
	\centering 
	\begin{tabular}{lcccc}
		\toprule  
		& Building 1 & Building 2 &Building 3&Building 4 \\
		\midrule
		Before EC(\$)&165.34& 38.06 &443.91&-203.69\\
		After EC (\$) &152.84 &35.58 &373.69&-315.79 \\
		Cost reduction (\$) &12.5 &2.48 &70.22&112.1\\
		\bottomrule 
	\end{tabular}
	\label{tabsocial}
\end{table}

\begin{figure}[!t]
	\centering
	\subfigure[]{	
		\includegraphics[width=0.23\textwidth,trim=0 0.2cm 0 0]{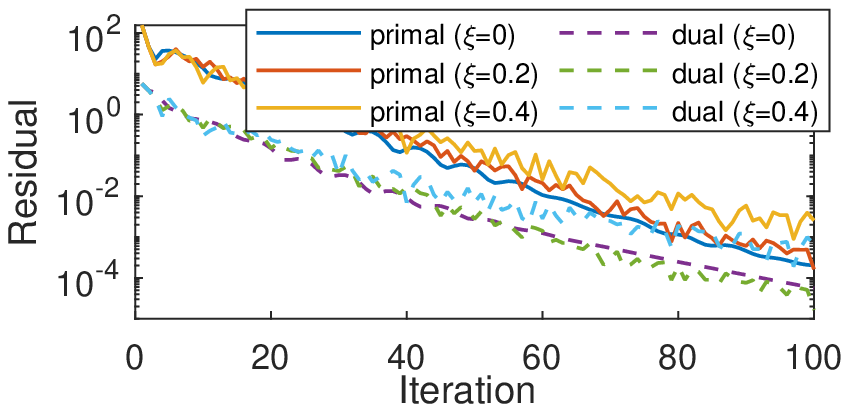}}
	\subfigure[]{
		\includegraphics[width=0.23\textwidth,trim=0 0.2cm 0 0cm]{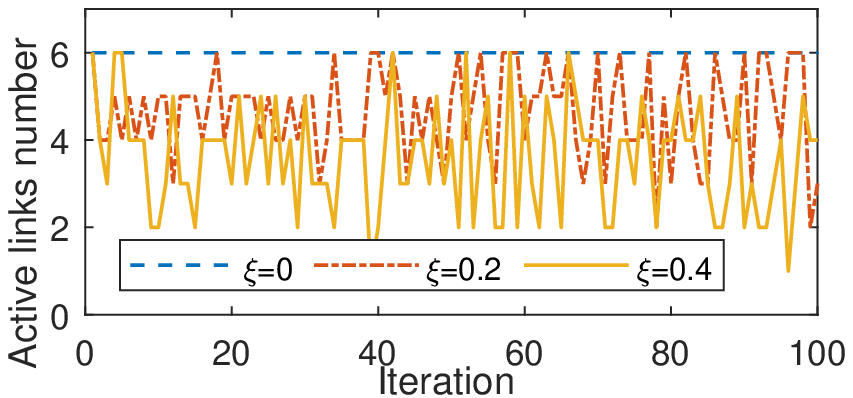}}	\caption{Convergence and communication condition curves.} 	\label{cvg}
\end{figure}

\paragraph{Convergence and communication in lossy networks}
Fig. \ref{cvg} (a) illustrates the residual curves of the proposed communication-failure-robust distributed algorithm in the lossy network ($\xi_{ij}$=$\xi$= 0.2, 0.4) and the ideal network ($\xi$=0). One can observe that, while the convergence speed slows down with increasing $\xi$, the proposed algorithm still converges to high-accuracy solutions compared to those in the ideal network.
Fig. \ref{cvg} (b) compares the number of active communication links during iterations, where only part of six links or even only one link is active. Note that all lines are equivalently enjoy the probability of $\xi$ to be inactive. These two observations demonstrate that the proposed algorithm is robust to random communication failures in the lossy network.

\section{Conclusion}\label{conclusion}
In this paper, we investigate the novel peer-to-peer energy cooperation framework for the building community. In the proposed framework, each building can directly trade with neighbors to collaboratively manage distributed energies. In addition, a communication-failure-robust distributed algorithm is proposed to preserve the agent privacy in presence of possible communication losses. Numerical results based on a four-building community demonstrate the effectiveness of the proposed framework in facilitating sustainable and harmonized energy use. Convergence analysis demonstrates that, the proposed algorithm is robust to communication losses in the network when only part of links (even one link) are active.


\bibliographystyle{IEEEtran}
\bibliography{pesgmbib}
\end{document}